\documentclass[aps,showkeys,superscriptaddress,preprint,tightenlines]{revtex4}
\usepackage{hyperref}
\usepackage{times}
\usepackage{amsmath}
\usepackage{graphicx}
\usepackage{epsfig}
\usepackage{dcolumn}
\usepackage{bm}
\usepackage{color}
\usepackage{longtable}
\usepackage{array}
\usepackage{multirow}

\newcommand{\jpsi}{J\kern-0.15em/\kern-0.15em\psi\kern0.15em}

\newcommand{\psip}{\psi(2S)}
\newcommand{\EE}{e^+e^-}

\newcommand{\beq}{\begin{equation}}
\newcommand{\eeq}{\end{equation}}
\newcommand{\bitm}{\begin{itemize}}
\newcommand{\eitm}{\end{itemize}}

\newcommand{\LHtwo}{LH$_2$}
\newcommand{\LDtwo}{LD$_2$}

\def\mycomm#1{\hfill\break\strut\kern-3em{\color{red}\tt ====> #1
\color{black}}\hfill\break}

\begin{document}

\title{\boldmath
Unraveling the Hyperon Puzzle in Neutron Stars 
\\
via Novel, High-Precision Hyperon Factories
}
\author{Chang-Zheng Yuan}
 \email{yuancz@ihep.ac.cn}
 \affiliation{Institute of High Energy Physics, Chinese Academy of Sciences,
 Beijing 100049, China}
 \affiliation{University of Chinese Academy of Sciences, Beijing 100049, China}
\author{Marek Karliner}
\email{marek@tauex.tau.ac.il}
\affiliation{School of Physics and Astronomy, Tel Aviv University, Tel Aviv 69978, Israel}

\begin{abstract}

The strong forces between nucleons ($N$=$p$, $n$) are fundamental to the visible universe. The interactions between hyperons (baryons with strange quarks) and nucleons are essential for the intrinsic properties of neutron stars.
Whereas the interactions between nucleons ($pp$, $pn$, $nn$) have been extensively studied, the interactions between nucleons and hyperons ($N\Lambda$, $N\Sigma$, $N\Xi$, $N\Omega$, ...) are not well understood, due to the small amount of relevant data, limited by the scarcity of suitable hyperon sources. 
Here we point out and investigate a new high-quality source: hyperons produced in $pp$ collisions, such as $pp\to pK^+\Lambda$, $pK\Sigma$, $pK\pi\Sigma$, $pKK\Xi$, $pKKK\Omega$.
At a fixed target experiment using proton beam with known momentum and liquid hydrogen target, $pp\to pK^+\Lambda$ can be produced copiously. By tagging $p$ and $K^+$, the flux and momentum of the $\Lambda$ can be determined precisely. By placing an additional target around the primary one, these $\Lambda$-s serve as an ideal source, enabling an unprecedentedly precise study of $\Lambda$ interactions with a wide range of targets. Similar methods can be used to obtain high-quality sources of other hyperons, such as $\Sigma$, $\Xi$ and $\Omega$.
These novel, high-statistics sources of hyperons with precisely known kinematics present new opportunities for applications in particle and nuclear physics, particularly in understanding the hyperon puzzle of neutron stars. 
We propose a new high-luminosity experiment with two nested concentric targets, optimized for such measurements. This concept can also be incorporated into existing experiments, such as HADES and CBM at FAIR, as well as proposed experiments, such as H-NS and HHaS at HIAF, by adding a second target without significant modification of the current detectors.

\end{abstract}

\keywords{hyperons, neutron stars, strong interaction, $pp$ collision}

\date{\today}

\maketitle


With core densities 5–10 times the nuclear saturation density ($\rho_0 \approx 0.16~{\rm fm}^{-3}$), neutron stars contain matter governed by strong interactions beyond the scope of perturbative quantum chromodynamics (QCD). In this environment, nucleons can transform into hyperons ($\Lambda$, $\Sigma$, $\Xi$, $\Omega$), deconfined quark matter, or exotic condensates. Their equation of state (EoS), which is constrained by astrophysical observations, directly encodes the nature of hyperon-nucleon (YN) and hyperon-hyperon (YY) interactions~\cite{Haidenbauer:2013oca,Vidana:2018bdi,Tolos:2020aln,Burgio:2021vgk,Chatziioannou:2024jsr}. 

Traditional models predict that the softening of the EoS due to presence of hyperons reduces the maximum mass of neutron stars to $<1.8M_\odot$. However, observations confirm the existence of millisecond pulsars with masses $\ge 2M_\odot$, such as PSR~J0348+0432~\cite{Antoniadis:2013pzd} and PSR~J0740+6620~\cite{NANOGrav:2019jur}.

Resolving this ``hyperon puzzle" requires a better understanding of the hyperon-nucleon interactions and the identification of repulsive mechanisms.
These include vector meson-mediated hyperon-hyperon interactions and hyperonic three-body forces which harden the EoS and constrain nonperturbative QCD and dense baryonic matter physics~\cite{Haidenbauer:2013oca,Vidana:2018bdi,Tolos:2020aln,Burgio:2021vgk,Chatziioannou:2024jsr}. 
All of this calls for high-precision data from hyperon-related processes, measured in scattering experiments.

Scattering experiments using various types of beams are essential for investigating fundamental interactions and the structure of matter at the subatomic level. Beams of long-lived charged particles and of photons are easy to produce, which is why many experiments using charged projectiles have been carried out for over 100 years since the trailblazing experiment shooting $\alpha$ particles into gold foil enabled Rutherford to infer the existence of the atomic nucleus~\cite{rutherford}. 
Since then, various beams, including $e^\pm$, $\mu^\pm$, $\pi^\pm$, $K^\pm$, protons, antiprotons, photons and heavy ions have been produced, enabling many scientific breakthroughs. Beams of some neutral particles, such as neutrons and neutrinos are relatively easy to produce, but difficult to control, i.e. have large momentum spread. Beams of other neutral particles, such as $K^0/\bar{K}^0$, and of long-lived hyperons ($\Lambda$, $\Sigma^{+,-}$, $\Xi^{0,-}$, and $\Omega^-$)\, have great physics potential, but are typically much more difficult to produce and control. 

Studies of hyperon-nucleon interactions began in the 1960s and have continued for over half a century~\cite{Engelmann:1966,SechiZorn:1969hk, Alexander:1969cx,Eisele:1971mk,Kadyk:1971tc,Hauptman:1977hr, Ahn:1997wa,Aoki:1998sv,Kondo:2000hn,Ahn:2005jz,J-PARCE40:2021bgw}. These studies employed $\pi^\pm$ or $K^\pm$ beams, with bubble chamber or scintillating fiber (SciFi) targets. These experiments have low statistics, typically observing a few tens to a few hundred events.

As discussed earlier, experiments with much higher statistics are clearly needed. In a previous article we proposed a new hyperon source, based on a completely different technique. We demonstrated that $\jpsi$ and $\psip$ produced in high-luminosity $\EE$ annihilation can be used as sources of large numbers of hyperons and antihyperons for novel nuclear and particle physics studies. These sources include hyperons $\Lambda$, $\Sigma^{+,-}$, $\Xi^{0,-}$, $\Omega^-$, as well as their antiparticles~\cite{Yuan:2021yks}.  
However, it will likely take a long time to realize this idea, due to the lengthy process of proposing, designing, and commissioning a super charm factory with sufficiently high luminosity~\cite{STCF,SCTF}. 
A proposal to use the BESIII detector with additional material between the beam pipe and the inner tracking detector CGEM was presented in Ref.~\cite{Zhang:2026lhm}. However, the improvement on the precision is limited by the insufficient luminosity of the BEPCII accelerator and the large radius of the beam pipe, as has been indicated in the BESIII measurements~\cite{Zhang:2025wlk}.

As shown below, in addition to the $\EE$ annihilation discussed above, a similar idea can be implemented much more quickly in fixed-target $pp$ experiments. This will enable us to obtain useful data quite soon.
We also point out that a wide range of novel high-precision physics measurements can be carried out by including an option in the detector design for inserting a variety of specific materials as a second target for the hyperons produced on the primary target. 
These measurements include hyperon-nucleon interactions~\cite{Haidenbauer:2013oca,Tolos:2020aln} and (multi-strange) hypernuclei~\cite{Gal:2016boi,Botta:2012xi,Pochodzalla:2016ncu,Burgio:2021vgk}. 

\begin{figure*}[htbp]
\centering
\includegraphics[width=0.8\textwidth]{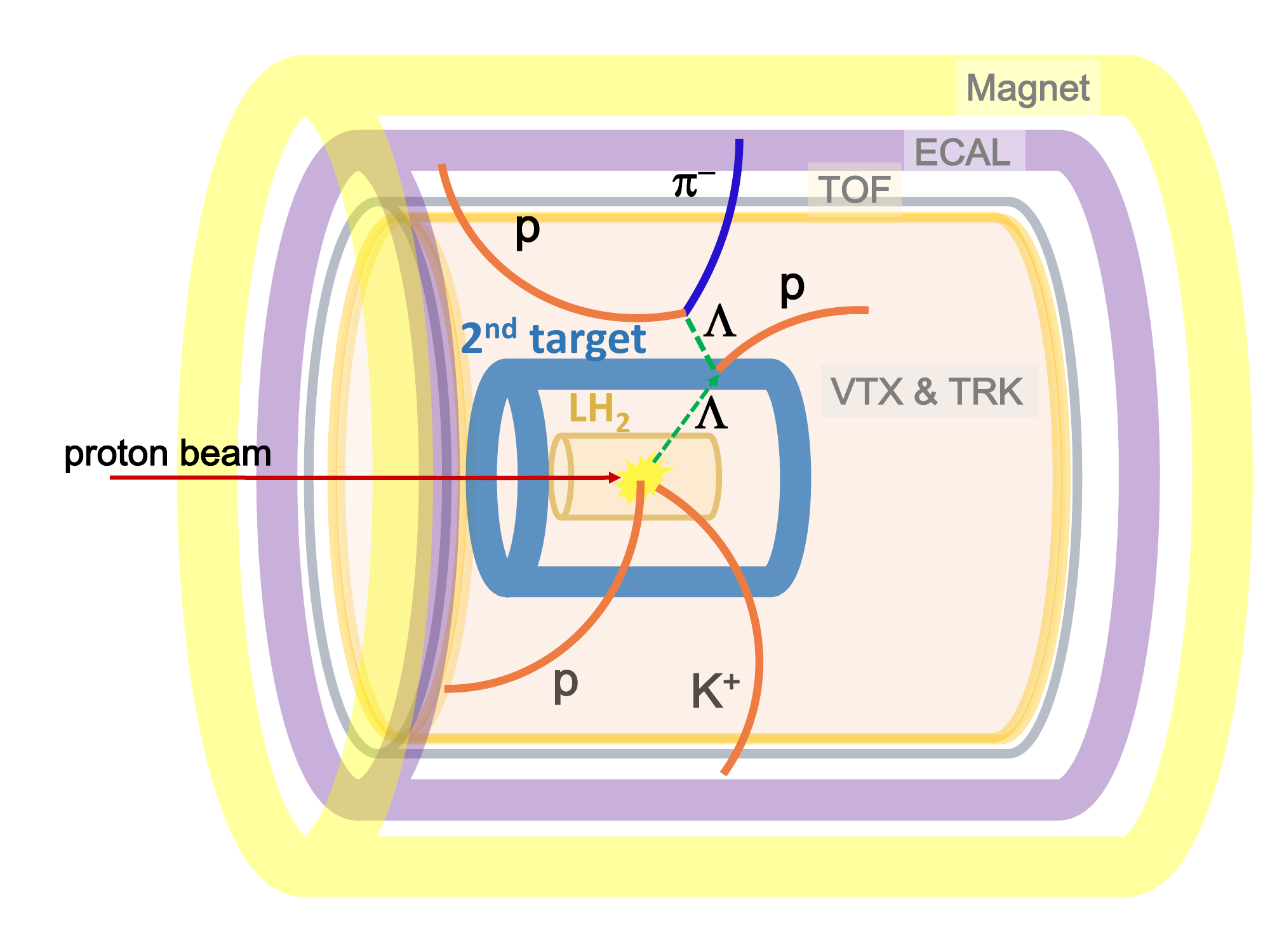}
\caption{\label{fig:two-target} 
A two-target experiment for hyperon production and high-precision measurements. The proton beam enters from the left and hits the \LHtwo\ primary target at the center of the detector. The other components, from the inside out, are the second target, the VTX~\&~TRK, the TOF, the ECAL and the magnet. The arcs and the green dashed lines denote the the following reaction products: $pp\to p K^+\Lambda$ in the \LHtwo\ target, $\Lambda p\to \Lambda p$ in the second target, and $\Lambda\to p\pi^-$ in the detector. }
\end{figure*}

Figure~\ref{fig:two-target} shows the conceptual design of a two-target experiment at a $pp$ collider. A conventional liquid hydrogen (\LHtwo) target is used to generate hyperons via $pp\to pK\Lambda$ or other processes, a second \LHtwo\ target (or solid target with free protons) or an \LDtwo\ target surrounding the first target serves as a proton or neutron target for the hyperons produced in the primary $pp$ collision. 
The detector includes a multi-wire gas chamber or a silicon-based detector for high-precision vertex and momentum measurements (VTX~\&~TRK), a scintillator detector or LGAD detector to measure the time-of-flight (TOF) for particle identification, and a crystal-based electromagnetic calorimeter (ECAL) for photon detection. The full detector is surrounded by a magnet. It is designed to precisely measure pions, kaons, protons, and photons. 

The second target is placed close to the primary target but separated by several to tens of millimeters. This allows the long-lived hyperons produced in $pp$ collision to travel to the second target and interact with the target materials.
The gap between the two targets ensures that the production and interaction vertices are well separated. This helps trigger and select signal events while suppressing backgrounds. The width of the gap is determined by the resolution of the vertex reconstruction. 
The hyperon produced at the primary target is tagged with the other particles produced in the $pp$ collision. Its momentum and direction are determined by the recoil of these other particles. The detector can fully reconstruct the final-state particles resulting from the hyperon-nucleon interaction in the second target. Conservation of four-momentum in the full interaction chain can be used to suppress background and improve resolution. 

As depicted in Fig.~\ref{fig:two-target}, the three-momentum vector of the tagged $\Lambda$ in the reaction $pp\to pK^+\Lambda$ is determined by measuring the recoil mass of the final-state $pK^+$, together with a static proton target and precise knowledge of the proton beam momentum. 
The tagged $\Lambda$ then interacts with a proton in the second target, producing a secondary proton and a $\Lambda$. The latter decays into a proton and a pion after traveling a short distance in the detector. 
In this case the reaction in the second target is $\Lambda p\to \Lambda p$. Knowing the number of tagged $\Lambda$s and the number of reconstructed $\Lambda p\to \Lambda p$ events, allows one to measure the total and the differential cross sections of this process. 

\begin{figure*}[htbp]
\centering
\includegraphics[width=0.8\textwidth]{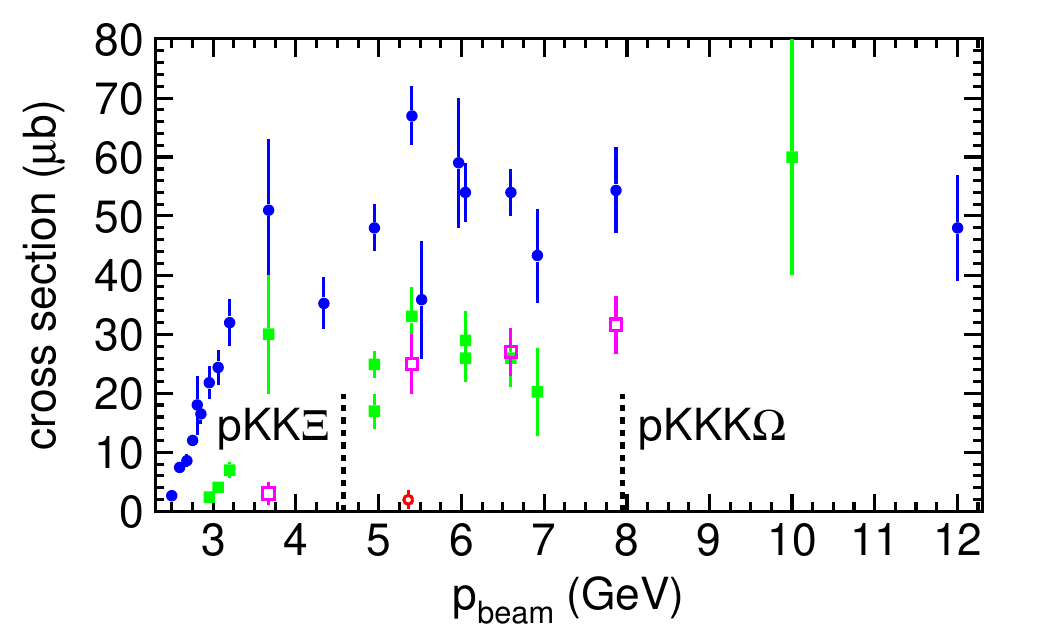}
\caption{\label{fig:cross} Hyperon production cross sections shown as a function of the beam momentum in fixed-target $pp$ experiments. The blue solid dots with error bars denote $pp\to p K^+ \Lambda$~\cite{Schopper:1980uer,Hofmann:1976rb, COSY-TOF:2010svd,COSY-TOF:2006tie,Bilger:1998jf,HADES:2014tlq}, the green solid boxes with error bars denote $pp\to p K^0 \Sigma^+$~\cite{Schopper:1980uer, TOF:2012tfr}, the pink open boxes with error bars denote $pp\to p K^+ \pi^+ \Sigma^-$~\cite{Schopper:1980uer}, and the open red dot with an error bar is an estimation of $pp\to p K^+ K^+ \Xi^-$~\cite{Agakishiev:2015xsa,HADES:2020pcx}, cf. text. The production thresholds for $pp\to p K^+ K^+ \Xi^-$ and $pp\to p K^+ K^+ K^0 \Omega^-$ are shown as black vertical lines.}
\end{figure*}

The position and thickness of the second target are determined by the event rate and vertex reconstruction resolutions. One would expect the vertex resolution to be sufficient to distinguish the final state particles from the $pp$ and hyperon-$p$ collisions. 
The best candidate materials for the second target are \LHtwo\ and \LDtwo, for measuring hyperon-$p$ and hyperon-$n$, respectively. However, solid targets such as polyethylene ($\rm (CH_2)_n$), deuterated polyethylene ($\rm (CD_2)_n$), lithium hydride ($\rm LiH$), or heavy water ($\rm D_2O$), could also be used.


In an experiment involving proton beam shot at an \LHtwo\ target, the momentum of the incident proton beam must be sufficient to produce one or more $s\bar{s}$ pairs, so that hyperon(s) are present in the final state.
Figure~\ref{fig:cross} shows the hyperon production thresholds in fixed-target $pp$ experiments for the simplest production processes. As can be seen, the $\Omega$ baryon with three strange quarks can be produced already at a beam momentum slightly below 8~GeV. 
Of course, the best energies must be found for producing one or more kinds of hyperons including the $\Lambda$, $\Sigma$, $\Xi$, and $\Omega$ baryons. This maximizes the sensitivity of tagging the hyperons and measuring their interactions in the second target.

The cross sections of hyperon production in $pp$ experiments, such as $pp\to pK^+\Lambda$, $pK^0\Sigma^+$, $pK^+\pi^+\Sigma^-$, and other associated- or multi-strangeness final states, have been measured in previous experiments, albeit with with substantial error bars.
Figure~\ref{fig:cross} compiles cross-section measurements of $pp\to pK^+\Lambda$ from Landolt-Boernstein~\cite{Schopper:1980uer}, CERN ISR~\cite{Hofmann:1976rb} and modern high-precision experiments COSY-TOF~\cite{COSY-TOF:2010svd,COSY-TOF:2006tie,Bilger:1998jf} and HADES~\cite{HADES:2014tlq}, and those of $pp\to pK^0\Sigma^+$ and $pp\to pK^+\pi^+\Sigma^-$ from Landolt-Boernstein~\cite{Schopper:1980uer} and COSY-TOF~\cite{TOF:2012tfr}.
These measurements show that single-strangeness production has cross sections on the order of a few microbarns near threshold. These cross sections increase with momentum, to a plateau of a few tens of $\mu$b above a beam momentum of 4~GeV.

In contrast, double-strangeness production, e.g., $\Xi^-$, is heavily suppressed, with cross sections around 1~$\mu$b ($(2.0\pm1.6)~\mu$b), as estimated from the measured ratio $\sigma_\Xi/(\sigma_\Lambda+\sigma_{\Sigma^0})$~\cite{Agakishiev:2015xsa,HADES:2020pcx}, shown as a red circle in Fig.~\ref{fig:cross}. 
There is no data for the triple-strange $\Omega^-$ channel. An educated guess is that production of $\Omega$ with three strange quarks is suppressed relative to $\Xi$ production with two strange quarks roughly as $\Xi$ production is suppressed vs. single-strangeness production, i.e. $\sigma_\Omega/\sigma_\Xi\sim \sigma_\Xi/(\sigma_\Lambda+\sigma_{\Sigma^0})$. Thus, one would expect the $\Omega$ production cross section to be at the ${\cal O}(20)$~nb level. 


The design of the targets depends on the required precision of the measurements. The proton beams expected in future accelerators will deliver approximately $10^{12}$ protons per second or more. With such a beam, the luminosity of an \LHtwo\ target that is 5~cm long is 
\[ {\cal L}=\phi\cdot \frac{\rho\cdot L}{m_p}\sim 2\times 10^{35}~{\rm cm}^{-2}s^{-1}=200~{\rm nb}^{-1}s^{-1}, \]
where the beam flux $\phi=10^{12}/s$, $\rho$ and $L$ are the density and length of the \LHtwo\ target, respectively, and $m_p$ the mass of proton.

Assuming the beam momentum is above the $\,pKK\Xi\,$ production threshold, the expected number of $\,pK\Lambda\,$ and $\,pKK\Xi\,$ produced during one month of operation (=$2.6\times 10^6$s) will be $2.6\times 10^{13}$ and $10^{12}$, respectively. During the same period, $10^{13}$ $\Sigma$ baryons will be produced. If the beam momentum is above the $pKKK\Omega$ threshold, a few billion $\Omega$s can be produced in one month.

Assuming the gap between the two targets is 1~cm and the second target is a few centimeters thick, say 5~cm, about 10\% of hyperons will reach and interact with the second target. This estimate also takes into account the geometric coverage of the second target and the momentum and angular distributions of the produced hyperons. 

Following the same analysis presented in Ref.~\cite{Zhang:2026lhm}, we estimate that the number of observed hyperon-proton interaction events will be ${\sim} 10^7$, ${\sim}10^6$, and ${\sim}10^4$ for $\Lambda$ ($\Sigma^\pm$), $\Xi$, and $\Omega$, respectively, in a process with a cross section of $1$~mb. Here we assume the detection efficiency is about 10\% for identifying $YN$ interactions. 
For a typical hyperon-nucleon cross section of a few to a few tens of millibarns, the sample size is large enough for a detailed analysis of the differential cross sections.


Proton beams for such a hyperon factory are already available at some laboratories. 
At FAIR in Germany, the SIS18/SIS100 synchrotrons supply proton beams for the HADES experiment, which is currently running, as well as for the proposed CBM experiment~\cite{Messchendorp:2025men}. The SIS18 accelerates protons up to kinetic energy of 4.5~GeV  (with a beam momentum of 5.36~GeV) with a maximum intensity of $2\times 10^{11}$ protons per cycle. The SIS100 is expected to be commissioned by the end of 2028. It will have proton beam momentum spanning a broad range--from 5 to 30~GeV--and a major increase in intensity of up to $2.5\times 10^{13}$ protons/cycle. This energy will be sufficient to produce all kinds of hyperons with open strangeness, $|S|=1$,~$2$,~$3$. 
The HIAF in China~\cite{Zhou:2022pxl} started operation at the end of 2025. It is designed to produce ion beams ranging from protons to uranium across a wide range of energies, at unprecedented intensities. For proton beams, the maximum momentum reaches 9.3~GeV, with an intensity of up to $6\times 10^{12}$ particles per second. 

The detector we propose can be placed in these beam lines to achieve the high-precision measurements discussed above. Experiments are already running or being designed with these beam lines. In principle, these experiments can be modified for our purposes without significantly affecting their original physics goals.

Two fixed-target experiments have been designed at FAIR: the HADES experiment at SIS18 and the CBM experiment at SIS100.

The HADES experiment at SIS18 is a versatile magnetic spectrometer designed to study dielectron production in reactions induced by pions, protons, deuterons, and heavy ions~\cite{HADES:2020pcx}. 
Its main features include a ring-imaging gas Cherenkov detector for electron–hadron discrimination, a tracking system composed of six superconducting coils that produce a toroidal magnetic field, drift chambers, and a time-of-flight array for further electron–hadron discrimination and a forward spectator detector. The \LHtwo\ target is a 5~cm long, 2.5~cm diameter cylindrical vessel. 
The maximum proton beam intensity allowed for current design of the experiment is $7.5\times 10^7$ protons per second. This results in a maximum average luminosity of $1.5\times 10^{31}~{\rm cm}^{-2}s^{-1}$. This is probably not sufficient for a high-precision measurement.
However, reducing the trigger rate could enable higher beam intensity during detector operation. This could be achieved by modifying the DAQ system or improving the trigger system.
One way to improve the trigger system would be to require energy conservation in both the primary and secondary targets.

The CBM experiment at SIS100 is designed as a high-rate, multipurpose fixed-target experiment capable of handling interaction rates of up to 10~MHz~\cite{Messchendorp:2025men}. It supports two detector configurations tailored to different physics cases.
One configuration includes a RICH and a transition radiation detector (TRD) for electron identification in hadron and dielectron measurements. The other configuration includes a muon chamber (MuCh) system for muon detection. Both configurations share a common tracking system and are selected based on the final states of interest, and both are suitable for the measurements of hyperon-nucleon interactions proposed above.
The SIS100 beam delivered to the CBM cavern reaches average intensities of up to $10^{12}$ protons per second. 
Using an \LHtwo\ target similar to the one used in the HADES experiment, a luminosity of $2\times 10^{35}~{\rm cm}^{-2}s^{-1}$ can be achieved. This is the exact case we discussed above.

Two fixed-target experiments are being designed at the high-energy nuclear physics terminal at HIAF: the Hyperon-Nucleon Spectrometer (H-NS) and the Huizhou Hadron Spectrometer (HHaS).

The H-NS detector is designed for high-precision measurement of hyperon and nucleon polarization~\cite{Lin:2025yzt}. Its design demands excellent momentum and spatial resolution, as well as a fast response time. The conceptual design comprises three main subsystems: a pixel silicon tracker for charged particle tracking, a Low-Gain Avalanche Detector (LGAD)-based Time-of-Flight (TOF) system for particle identification, and a electromagnetic calorimeter (ECal) for neutral particle detection.

The HHaS aims to revolutionize the fields of particle and nuclear physics by achieving an unprecedented increase in the statistical precision in the study of light hadrons, by $\sim$2 orders of magnitude, compared to existing experimental setups (e.g., KLOE-2 and BESIII)~\cite{Chen:2025ppt}. The HHaS detector system comprises a solenoidal magnet, a pixel tracker, a time-of-flight (TOF) detector and an electromagnetic calorimeter (EMC).

The targets at H-NS and HHaS have not yet been designed. With an \LHtwo\ target similar to that used in the HADES experiment and a beam flux of $6\times 10^{12}/$s from the accelerator, a maximum luminosity of $12\times 10^{35}~{\rm cm}^{-2}s^{-1}$ can be reached. The design of the aforementioned two detectors meets the requirements of our proposal. The H-NS design, which is optimized for hyperon polarization measurements, can potentially also study the hyperon production modes proposed above. The detector has excellent performance in vertex reconstruction and momentum measurement. The HHaS can also trigger the hyperon events without affecting the original design. 
With a second target included in the design of these experiments, physics goals of both the original experiments and the high precision hyperon-nucleon interaction experiment can be reached. 


In summary, we have demonstrated that a fixed-target $pp$ experiment can produce a large number of hyperons whose kinematics can be precisely determined. This opens up prospects for novel studies particle and nuclear and particle physics. 
The produced particles include all the long-lived hyperons $\Lambda$, $\Sigma^{+,-}$, $\Xi^{0,-}$, and $\Omega^-$. By placing a second target around the primary one, a variety of particle and nuclear physics experiments can be performed, especially high-precision measurements of hyperon-nucleon interactions. 

This idea can be implemented in existing and proposed experiments, by adding
a second target without significantly altering the current designs.

Traditional setups require producing many different kinds of beams for various experiments and sharing accelerator time among them. This requires substantial resources in terms of manpower and funding, hindering such experiments. In contrast, the proposed approach enables experiments with different hyperon beams to be conducted simultaneously, requiring no additional infrastructure and minimal further investment.

The new high-intensity sources of hyperons, with precisely determined kinematics, offer fresh opportunities to advance our understanding of the hyperon puzzle in neutron stars. 

\section*{Acknowledgments}

This work is supported in part by National Natural Science Foundation of China (NSFC) under contract No.~12361141819, the ISF-NSFC joint research program (grant No. 3166/23), as well as the Munich Institute for Astro-, Particle and BioPhysics (MIAPbP) which is funded by the Deutsche Forschungsgemeinschaft (DFG, German Research Foundation) under Germany's Excellence Strategy – EXC-2094 – 390783311. We thank Ai-Qiang Guo, Xiao-Long Wang, and Zhao-Ling Zhang for discussions during preparation of this article. 


\end{document}